\documentclass[prb,preprint]{revtex4-1}

\usepackage{amsmath}  
\usepackage{amsfonts} 
\usepackage{graphicx} 

\begin{document}

\title{Measurements and analysis of current--voltage characteristic of a \emph{pn} diode for an undergraduate physics laboratory}

\author{Enrico Cataldo}
\email{enrico.cataldo@unipi.it}
\affiliation{Dipartimento di Fisica ``E. Fermi'', Universit\`a di Pisa, Pisa, 56126 Italy}

\author{Alberto Di Lieto}
\email{alberto.dilieto@unipi.it}
\affiliation{Dipartimento di Fisica ``E. Fermi'', Universit\`a di Pisa, Pisa, 56126 Italy}

\author{Francesco Maccarrone}
\email{francesco.maccarrone@unipi.it}
\affiliation{Dipartimento di Fisica ``E. Fermi'', Universit\`a di Pisa, Pisa, 56126 Italy}

\author{Giampiero Paffuti}
\email{giampiero.paffuti@unipi.it}
\affiliation{Dipartimento di Fisica ``E. Fermi'', Universit\`a di Pisa, Pisa, 56126 Italy}

\date{\today}

\begin{abstract}
We show that in a simple experiment at undergraduate level, suitable to be performed in classes of science and engineering students, it is possible to test accurately, on a popular 1N4148 \emph{p-n} diode, the range of the junction currents where the Shockley equation model can be considered satisfactory. The experiment benefits from a system of temperature control and data collection driven in a LabVIEW environment. With these tools a large quantity of data can be recorded in the temporal frame of a lab session. Significant deviations of the experimental \emph{I--V} with respect to the ideal behaviour curve predicted by the Shockley equation are observed, both at low and high current. A better agreement over the entire range is obtained introducing, as is customary, a four parameters model, including a parallel and a series resistance. A new iterative fitting procedure is presented which treats the \emph{I--V} data of different regimes on the same level, and allows a simultaneous determination of the four parameters for each temperature selected. Moreover, the knowledge of the temperature dependence of saturation current is used to estimate the energy gap of silicon. The connection of a macroscopic measure with a microscopic quantity is another valuable feature of this experiment, from an educational point of view.
\end{abstract}


\maketitle %

\vspace{2pc}


\section{Introduction}

Current-Voltage (\emph{I--V}) curves are routinely used  to give synthetic graphical information of non-ohmic electronic devices such as semiconductor diodes. Moreover, a lot of information about the properties of conducting materials and devices is obtained through the analysis of the current-voltage dependence both in DC and in AC regimes. In the literature there is a large amount of \emph{I--V} based studies for a wide variety of devices, encompassing homogeneous, composite materials and even biological structures.

It is worth mentioning that \emph{I--V} characteristic measurements are used for the assessment of the performances of conventional photovoltaic cells\cite{Zhang} and non conventional photodevices such as a single carbon nanotube exhibiting the peculiar behaviour of a photodiode\cite{Barkelid}; in the investigation of the collective conduction of charge observed in suitable materials and originating in the formation, through a quantum mechanism, of a coherent travelling charge wave\cite{Thorne}; in electrophysiology\cite{Johnston}; in the study of the properties of electrical conduction of a few or even single biomolecules such as short strings of DNA adherent across the segments of conducting surfaces\cite{Fink}. We limit ourselves to cite only these few examples, with the aim to stress the importance and spreading of the \emph{I--V} characteristic tool in many different fields, and to support the call of a wide use of this technique in the education of scientists and engineers toward experimental practice.

The \emph{p-n} semiconductor junction, one of the simplest elements with a non linear response, is still the first choice starting point in the study of semiconductor devices which is a primary goal of modern labs for science and engineering undergraduate students\cite{Meehan}. In Ref. \cite{Neudeck} the introduction states: \emph{``The \emph{p-n} junction diode is the most fundamental of all the semiconductor devices (...). So basic is the theory of operation that many engineers have stated that to understand the \emph{p-n} junction qualitatively and quantitatively helps one to understand the majority of all solid state devices.''}

In educational literature there are several examples of experimental papers studying the \emph{p-n} junction both in diode devices or in transistors\cite{Inman}, focusing on the conformity of the measured \emph{I--V}  characteristics with the modelling equations\cite{Martil} or on particular aspects, such as the possibility to measure the energy band-gap in diodes\cite{Canivez} or in light emitting diodes\cite{Precker}.

In this paper we present an apparatus, used in a class fractionated into small groups of two-three students, to investigate the \emph{I--V} characteristics of a commercial diode at controlled temperatures, in order to test the compliance of the well-known Shockley equation with the behaviour of a real device. Temperature control, data collection and analysis are directed into a LabVIEW environment.

We present a detailed data analysis in order to invite students to appreciate the caution necessary to avoid some common pitfalls of the fitting procedures and to better appreciate the meaning and limits of a physical model. We also show how a deeper insight into the physical basis of the \emph{p-n} junction could be obtained by examining the dependence of the basic diode model parameters from surroundings temperature. As an example, we report the exponential dependence of the saturation current on the inverse of the temperature and we show how the constants involved give information on the barrier height of the junction.

\section{Terms of the model}

The main objective of a laboratory session dedicated to the rectifying diode consists in the evaluation of the according degree of the measured curve with the models.
The ideal \emph{I--V} characteristic of a \emph{p-n} junction is not commonly treated in a general physics textbooks, and it is usually approximated by the popular Shockley equation, omitting its considerably complex theoretical origin\cite{Neudeck}:
\begin{equation}
\label{eq1}
I = I_S\left[\exp{\left(\frac{V}{nV_T}\right)}-1\right]
\end{equation}
The properly called Shockley equation has the \emph{non ideality factor} $n=1$, and is appropriate only for ideal junctions. For real junctions, $n$ is greater than 1.
The \emph{thermal voltage} $V_T = kT/e$  has a value of 25.8 mV at room temperature.
The saturation current $I_S$ describes the level of conductivity of the diode, and ideally it coincides with the value of the current when a large reverse voltage is applied to the junction. It depends (in a very complex way) on the constructive parameters of the diode, such as the area and the depth of the junction, or the doping spatial uniformity.
It is possible to reasonably express its main dependence on the temperature following the argument that $I_S$ grows with the population of the conducting level of the semiconductor carriers and that this follows a stationary thermodynamic distribution, so that:
\begin{equation}
I_S = I_0\exp{\left(-\frac{E_G}{kT}\right)}
\label{issempl}
\end{equation}
where $E_G$ is the energy gap of the bulk semiconductor.\\
Using a similar argument adopted normally to introduce the non-ideality factor $n$ in the Shockley equation, the same factor must be taken into account in the exponential dependence of $I_S$:
\begin{equation}
I_S = I_0\exp{\left(-\frac{E_G}{nkT}\right)}
\label{isconn}
\end{equation}

Although in the real diode the value of $n$ is only approximately constant with V, in a quite large interval of direct currents eq. (\ref{eq1}) reproduce well the \emph{I--V} curve with constant values of $I_S$ and $n$, and these two parameters can be easily determined  with some fitting procedure, even graphical. In fact, when $V \gg nV_T$, the logarithmic graph of the current show a clear initial linear trend.
The dependence of the fitted value of $n$ on the portion of the characteristic curve analyzed has already been  discussed previously\cite{Chand}.

To increase the quality of the fit of experimental data, some authors \cite{Kami} proposed junction models consisting of a sum of exponentials, each with different $I_S$ and $n$ parameters. Each contribution is considered valid in different regions of the \emph{I--V} curve and some kind of connection in the contiguous regions is required. The weakness of these strategies is the lack of an underlying simple physical picture and the increase of the number of model parameters.

Finally, it is easy to observe significant deviation from the behaviour predicted by the model of eq. (\ref{eq1}) on both sides, of large and very low currents. In the first case the observed direct current is lower than that predicted by the model, and the exponential trend is weakened so that the logarithmic graph stays under the straight line calculated in the range of intermediate current. At the lowest current a larger direct current is observed with respect to that calculated with the parameter estimated in the region of intermediate currents.
The origin of the large current deviation could be justified at an elementary level as the lowering of the voltage difference across the junction with respect to the applied \emph{fem} due to ohmic voltage drop in the semiconductor bulk hosting the junction. On the other hand, the low current regime originates from the presence of alternative paths for the current. When the junction current is very low, these parallel paths give a comparable contribution to the total current with a different dependence on $V$ which is rapidly overcome from the exponential term of the junction.
\begin{figure}[!ht]
\centering
\includegraphics{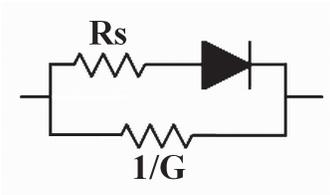}
\caption{Electrical circuit corresponding to the four parameters model of the real diode.}
\label{modello}
\end{figure}
Then, a more accurate model for the stationary \emph{I--V} characteristic of the diode can be formulated introducing a series resistance $R_s$, accounting for the ohmic voltage drop across the semiconductor bulk, and a parallel resistance $R_p = 1/G$, accounting for alternative paths: Fig. (\ref{modello}) shows the equivalent electrical circuit; this model has four parameters, and eq. (\ref{eq1}) modifies into:
\begin{equation}
\label{model2}
I = I_p + I_J = GV + I_S\left[\exp{B\left(V-R_sI_J \right)}-1\right]
\end{equation}
Where $I_J$ is the current flowing into the junction, and $I_p = GV$ is the current flowing into the parallel resistance; a similar modeling has been proposed in ref. \cite{Mika}.
Parallel and series resistances are normally introduced in the modelization of the photovoltaic cells and in the evaluation of its performance, so that the comparison of this model to the experimental data of a diode is also valuable as an introduction to the parametrization of the photovoltaic devices. A good discussion of the state of the art of the applied numerical methods is found in a recent paper\cite{Hansen}.

\section{Experimental}
\begin{figure}[!ht]
\centering
\includegraphics{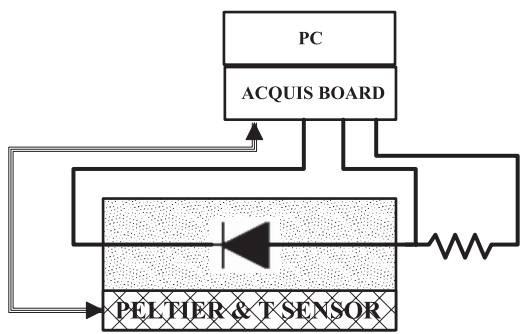}
\caption{Schematic view of the experimental setup.}
\label{setup}
\end{figure}

A small (4 $\times$ 4 $\times$ 4 cm$^3$) brass cube is placed between two 30 W thermoelectric power sources/sinks assembled with commercial Peltier modules. It hosts both the device being tested, a commercial 1N4148 silicon \emph{p-n} diode, and a calibrated thermistor. The direction of the heat flux can be selected through the polarity of the electric power supply connected to the modules. The heat exchanged is regulated through the duration of the power pulse in proportion to the difference between the requested temperature and the temperature registered by the thermistor. The system stabilizes the temperature of the central volume of the cube in a few tens of seconds to the value set by the operator in the range 10-100 $^\circ$C with an accuracy better than 1 $^\circ$C. The \emph{I--V}  characteristic is then measured with two differential channels of a National Instruments NI6221 data acquisition board equipped with multiplexed 16-bit ADC with selectable bipolar full scales. One of the ADC channels samples the voltage $V$ across the electrodes of the diode with a resolution of 0.3 mV and the other channel senses the current $I$ crossing the diode by  measuring the potential drop on a calibrated resistor $R$ in series with the diode, with a nominal resolution of 0.02 $\mu$A. $V$ is varied applying to the series diode-$R$ the variable signal of a digital-to-analog channel (DAC) of the NI6221 board. The control of the thermal cycle and of the data taking is programmed devising a LabVIEW virtual instrument. Initially, a set of temperatures is selected by the operator. After this, the controlled thermal cycles operate until the value of each $T$ is stable into a few tenths of $^\circ$C. The couples [V,I] are saved in a file and are labelled with the temperature measured just before and just after the voltage scan. The difference of two readings of the temperature is ever less than 0.5 $^\circ$C in absolute value, the sign being dependent on the direction of the last heat flux cycle. Thereby, we can evaluate that the value of the diode temperature is known with a precision of 0.2 $^\circ$C on average.

\section{Results and Discussion}
\subsection{Shockley model}

\begin{figure}[!ht]
\centering
\includegraphics{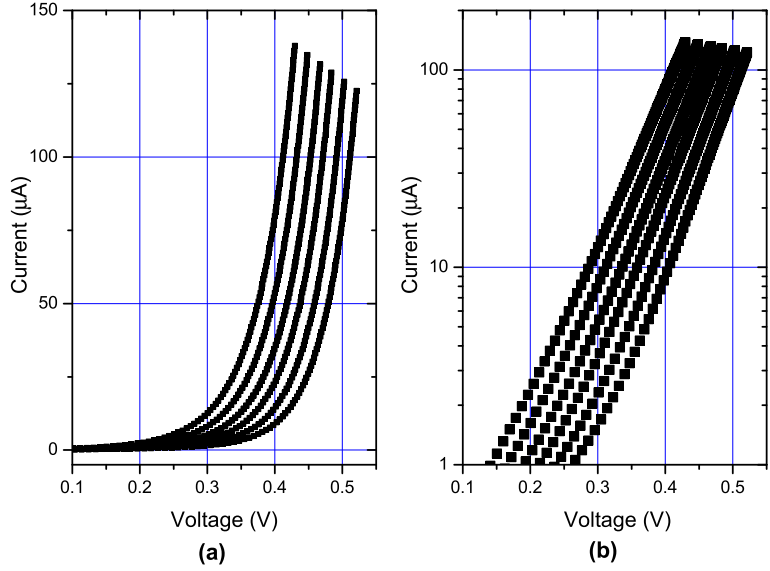}
\caption{I-V pairs taken at six different temperatures (20.7 , 29.7, 39.3, 47.8, 57.2 and 66.3 $^\circ$C, from right to left), in linear scale (a) and in $\log_{10}$ scale (b) of the currents; it is evident the alignment of the experimental points, suggesting the exponential dependence of $I$ on $V$.}
\label{fig1}
\end{figure}

Fig. \ref{fig1} shows a set of \emph{I--V} curves taken at different temperatures, plotted both on linear and logarithmic scale: it is evident in the right panel the alignment of the experimental points, which is so much better for the higher temperatures and at the higher value of $V$. The entire set of data required a time of less than ten minutes so that the measurements could be made in a standard lab session of 2-3 hours leaving time for a preliminary data analysis and a possible reiteration of the data collection.

The linearity of the experimental \emph{I--V} characteristic in the logarithmic current scale reported in Fig. \ref{fig1}b indicates, at a glance, the region where an exponential dependence of $I$ on $V$ is likely.
A first analysis of data can be done by using eq. (\ref{eq1}), limiting the measurements to the region of moderate injection current where $I$ is dominated by diffusion. Consequently, we can neglect non-linear effects on the current, reducing the number of parameters to the two appearing in eq. (\ref{eq1}).
\begin{figure}[!ht]
\centering
\includegraphics{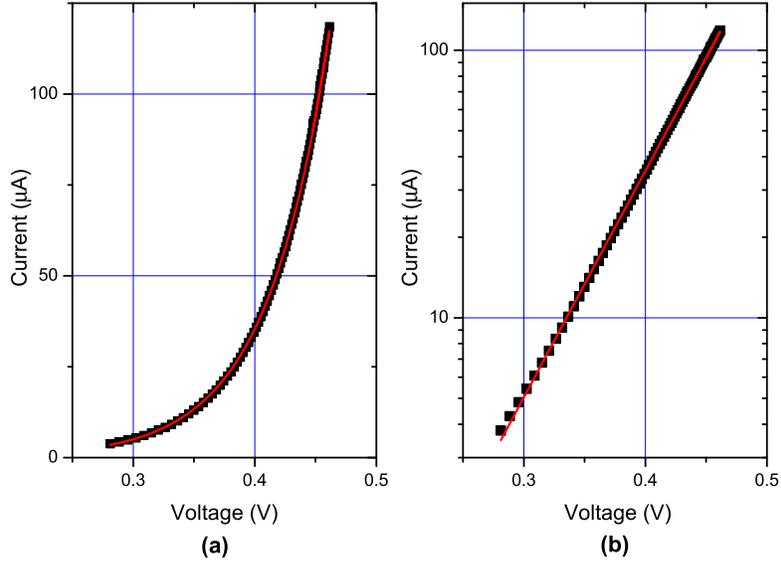}
\caption{Exponential fit of data corresponding to the temperature T= 47.8$\pm$0.2 $^\circ$C in linear scale (a) and semi-logarithmic scale (b). Normalized $\chi^2$ is of the order of unity for uncertainties of the current of the order of 0.05 $\mu A$ which is three times larger than the instrumental resolution and can be attributed to the experimental variability.}
\label{fig2}
\end{figure}
We write the eq. (\ref{eq1}) in the more compact form:
$$I = A \, [\exp{(BV)}-1]$$
where $B=e/nkT$. Fig. \ref{fig2}a shows a non linear fit of this equation, through a standard least square routine\cite{Matlab}, which gives a value of $B=e/nkT=19.48\pm 0.03\, V^{-1}$, largely different from the ideal diode constant at the same temperature $T = 47.8\pm 0.2 ^\circ C$ - $B_{id} = e/kT=36.20 \, V^{-1}$. The agreement of the model with the data requires a value:
$$ n = \frac{B_{id}}{B} = 1.86\pm 0.01$$
which is lower than the value of the SPICE models for 1N4148\cite{NXP} diode but agrees with other experiments \cite{Su}, pointing out the variability of this parameter due to constructive tolerance or data analysis.

A different treatment of data, often encountered in literature, considers the large value of the product $BV$ in the range investigated. In this case, one can neglect the term -1 in the parenthesis and can take the natural logarithm of the measured current, finding:
$$ \log{I} = \log{A}+BV$$
The linear fit of the logarithm of the data gives $\log{A} = -4.29\pm 0.02$ with the current $A$ expressed in $\mu$A and $B=19.63\pm 0.04\, V^{-1}$ (see Fig. \ref{fig2}b). The differences between these values and those obtained  with the complete equation are rather small, confirming the feasibility of this common and straightforward processing of data, even if a simple analysis of residuals of the two models shows that the linearized one is less accurate in the full range investigated.

\subsection{Deviations from the Shockley model}
\begin{figure}[!ht]
\centering
\includegraphics{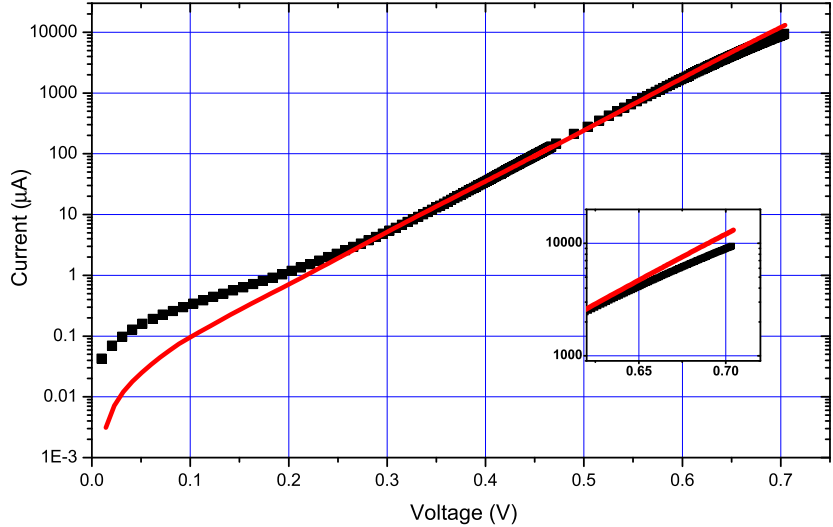}
\caption{\emph{I--V}  data at $T= 47.8\, ^\circ C$ plotted in semi-logarithmic scale. The continuous line is the best fit curve obtained with the Shockley model with two parameters applied to the data with $2\, \mu A < I < 120\, \mu A$. The inset is a zoom of the high current data. It is evident that at very low currents ($I <1\, \mu A$) and at high current ($I > 1\, mA$) the \emph{minimal} model is not accurate.}
\label{figura3}
\end{figure}
Fig.(\ref{figura3}) shows the plot of $\log(I)$ \emph{vs.} $V$ for a larger range of $V$. The curve superimposed to the experimental data is the best fit obtained with the eq. (\ref{eq1}), and it is evident that the two parameters model systematically underestimates lower values of the current and overestimates its higher values. A more realistic comprehension of the nature of the different contributions to the diode current, therefore, requires  a study of these limiting regimes.

As explained above, besides the two parameters of the Shockley equation $I_S$ and $n$, two supplementary parameters must be taken into account: $G$ and $R_s$. \\
If the series resistance is neglected, eq. (\ref{model2}) simplifies into:
\begin{equation}
\label{3parlow}
 I = GV + I_S\left(\exp{BV}-1\right)
\end{equation}
with $I$ and $V$ being the current and the voltage sensed at the diode leads, as before.

In ref. \cite{Mika} the authors declare the unsuitability of the analysis of the forward \emph{I--V} curve to find $G$ and suggest the extraction of this parameter through a graphical treatment of the reverse biased diode characteristic. Indeed, for $V << nV_T$ the linear approximation of the exponential brings a linear relation between the current and the voltage:
\begin{equation}
\label{nose}
    I = (G+BI_S)V
\end{equation}
which is not useful to calculate $G$ unless $I_S$ and $B$ are known or if the second term of the function of eq. (\ref{nose}) is negligible, as in this case ($B I_S \simeq 0.2 \mu S$). As a matter of fact, we find it suitable to apply the same least square nonlinear fitting routine using eq. (\ref{3parlow}) as the model equation. As the first term $GV$ gives an appreciable contribution only at very low values of $V$ we limit the experimental data up to a guess threshold current $I_0$. The value of the small signal conductance $G$ can be determined executing the corresponding numerical fit with three parameters, $I_s$, $B$ and $G$.

\begin{figure}[!ht]
\centering
\includegraphics{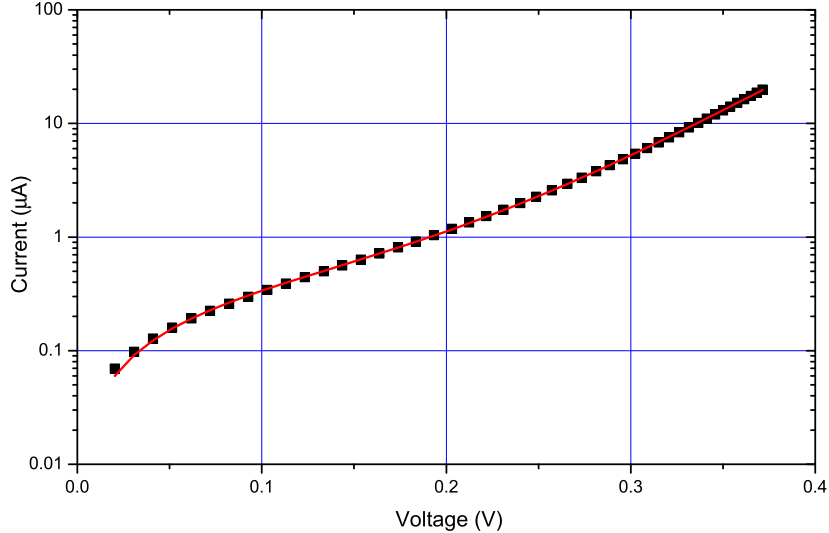}
\caption{\emph{I--V}  data corresponding to the temperature $T= 47.8\, ^\circ C$ limited to $I < 20\, \mu A$. The continuous line is the result of the fit of experimental data (black square) with the model of eq. (\ref{3parlow}). The fitted parameters are $Is = 9.72\pm 0.17\, nA$ and $B = q/nkT = 20.30\pm 0.05 \,V^{-1}$ and $G= 2.75 \pm 0.01\, \mu S$.}
\label{figura4}
\end{figure}
The result of the fitting procedure are shown graphically in Fig.~(\ref{figura4}) and the found values for parameters are reported in the caption. It is worth noting that $I_S$ and $B$ values are not in accordance with those extracted with the minimal model in the central range of the currents. This confirms that the calculated ideality factor $n$ depends on the part of the \emph{I--V} curve used in its evaluation.

\begin{figure}[!ht]
\centering
\includegraphics{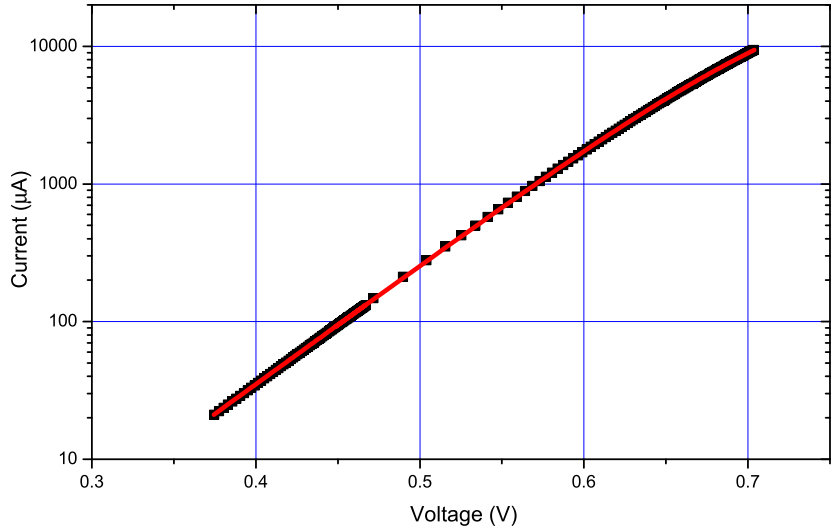}
\caption{\emph{I--V}  data corresponding to the temperature $T= 47.8\, ^\circ C$ selected with $I > 20\, \mu A$. The continuous line is the result of the fit of experimental data (black square) with the model of eq. (\ref{3parhi}). The fitted parameters are $Is = 13.4\pm 0.12\, nA$ and $B = 19.78\pm 0.02 \, V^{-1}$ and $R_s= 2.32 \pm 0.03\, \Omega$.}
\label{figura5}
\end{figure}

At the higher values of the diode current, the series resistance $R_s$ becomes important, while the admittance $G$ gives a negligible contribution.
We may then rewrite eq. (\ref{model2}) as:
\begin{equation}
\label{3parhi}
V(I) = \frac{1}{B}\log\left(\frac{I }{I_S}+1\right) + R_s I
\end{equation}
and it can be used as a fit model for the \emph{I--V} data at higher currents, giving the best values of the three parameters $I_S$, $B$ and $R_s$.
The result is shown in Fig. (\ref{figura5}), and the found values for parameters are reported in the caption.

The values of the parameter $B$ are approximately coincident, by few percent, in the two parts of the curve and the coincidence is maintened by varying the value of $I_0$. The values of the saturation currents differ by the order of ten percent, and the agreement is better for higher temperatures.
The value of $R_s$ is comparable with other measurements found in literature \cite{Omar}, although there the data is rather different from different sources (e.g. \cite{NXP}). We were not able to find a reliable value of $G$ for the 1N4148 diode in the literature.

In conclusion, the two sets of parameters obtained by fitting data with model eq. (\ref{3parlow}) for the region $I<I_0$ and with model eq. (\ref{3parhi}) for the region $I>I_0$, where $I_0$ is of the order of tens of $\mu$A, show some discrepancies, so a better analysis for the complete range of current is needed.

\subsection{Single iterative fitting procedure}
The analysis presented in the previous paragraph clearly shows that the first term of eq. \ref{model2} is significant only in the very low current region (see Fig. (\ref{figura4})), while at higher currents the exponential term dominates and the second term alone is able to describe  the experimental data well (see Fig. (\ref{figura5})). This physical argument is the starting point of a numerical procedure which allows us to consider all the data together, using an iterative calculation.

We point out that by substituting $I_J = I - GV$ in the argument of the exponential of eq. \ref{model2}, an equation $I = f(I,V)$ could be written as the objective function of a fitting procedure suitable for models expressed in an implicit form. Nevertheless, this kind of algorithms is very complex and frequently suffer from numerical instabilities.

Here we describe an alternative method, based on an iterative fitting procedure for models expressed in an explicit form, leading to the simultaneous determination of four parameters necessary to reproduce the DC diode characteristic in a large region of currents.
This method can be summarized in a few steps, including the iteration described in the loop section, where $i$ is the flowing index, $I$, $I^i$ and $I_J^i$ are respectively the experiment values, computed values and computed junction values of the current:
\begin{itemize}
\item[\textbf{1}] The first few data at very low currents are used to compute a first guess $G^i$ ($i=0$) of the parallel conductance with a linear fit of the model $I = GV$.
\item[\textbf{2}] \textbf{ BEGIN LOOP} Computing $I_J^i = I - G^i V$ gives the estimate of the junction current.
\item[\textbf{3}] The explicit model $V=V(I_J)$ of eq. (\ref{3parhi}) is fitted with the data calculated at step 2 and the estimate of $I_s^i$, $B^i$ and $R_s^i$ is found.
\item[\textbf{4}] Updated values of the junction current $I_J^{i+1}$ are found solving numerically eq. (\ref{3parhi}) with the measured $V$ and assuming the estimate of the three parameters $I_S^i$, $B^i$ and $R_s^i$.
\item[\textbf{5}] Updated values of total current $I^{i+1}$ are computed as $I^{i+1} = G^i V + I_J^{i+1}$.
\item[\textbf{6}] The first few data of $I^{i+1}$ are compared to $I$ to obtain the new value $G^{i+1}$:
     $I - I^{i+1} = \delta G V$, $G^{i+1} = G^i + \delta G $.
\item[\textbf{8}] \textbf{END LOOP} The four parameters with index $i+1$ are compared to previous values: if the change is much less than the estimate of their errors the loop ends, otherwise it continues, returning to step 2.
\end{itemize}

\begin{figure}[!ht]
\centering
\includegraphics{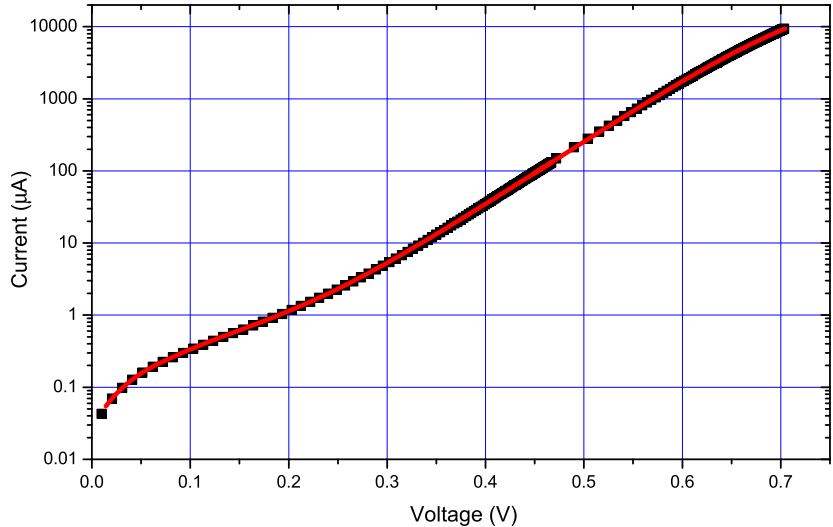}
\caption{\emph{I--V}  data corresponding to the temperature $T= 47.8\, ^\circ C$. The line is the result of the iterative fitting procedure described in the text. The agreement is good in the full range of the currents considered with $Is = 10.5\pm 0.2\,$, $B =20.2 \pm 0.1 V^{-1}$, $R_s= 2.75 \pm 0.05\, \Omega$ and $G = 2.80 \pm 0.05\, \mu Si$.}
\label{figfit}
\end{figure}

After a few iterations the four parameters converge toward stable values and the procedure may be stopped. The final result is shown in the Fig.~(\ref{figfit}) where the agreement of the model with the experimental data is very good over the entire range.

Table \ref{parmod} shows the extracted parameters with the different models presented, using data of a specific current region when needed. The last set is obtained by using the iterative fitting procedure.

\begin{table*}[!ht]
\caption{Experimental parameters of the models for the \emph{I--V}  data at $T= 47.8\, ^\circ C$}
\label{parmod}
\centering
\begin{tabular}{|c|c|c|c|c|c|c|}
\hline
\bfseries &\bfseries G ($\mu$S)& \bfseries $R_s (\Omega)$& \bfseries $I_S$ (nA)& \bfseries B (V$^{-1}$) & \bfseries n \\
\hline
Minimal model                          & -           & -          & 14.63 (0.16) & 19.48 (0.03) & 1.856 \\
Low $I$ (eq. (\ref{3parlow}))          & 2.75 (0.01) & -          & 9.72 (0.17) & 20.30 (0.05)  & 1.783  \\
High $I$ (eq. (\ref{3parhi}))          & -           & 2.32 (0.03)& 13.4 (0.12)  & 19.78 (0.02) & 1.830  \\
Full curve (eq. (\ref{model2}))        & 2.80 (0.05) & 2.75 (0.05)& 10.5 (0.2)   & 20.2 (0.1)   & 1.790  \\
\hline
\end{tabular}
\end{table*}

\subsection{Determination of energy band gap from temperature dependence of parameters}

In this section we extract energy band gap from temperature dependence of $I_S$ parameter, fitting \emph{I--V} data taken at temperature between 10 and 100 $^\circ$C.
$E_G$ is not explicitly present in the Shockley equation as it is enclosed only in the saturation current $I_S$; a more detailed expression of eq. (\ref{isconn}) is given by:
\begin{equation}
I_S = A T^2\exp{\left(-\frac{E_G}{nkT}\right)}
\label{Richa}
\end{equation}
where $A$ depends on the geometry of the junction and the doping densities in the device.

There are two counteracting effects of an increase of $n$ on the diode current. An increase of $n$ causes a decrease of the current $I$ in eq. (\ref{eq1}) the other parameters being fixed but, on the other hand, the same increase determines an increase of $I_S$, at a given temperature. Because of the larger value of $E_G$ with respect to the voltage across the junction, the effect on $I_S$ prevails and at fixed parameters, an increase of $n$ entails an increase of $I$.

\begin{figure}[!ht]
\centering
\includegraphics{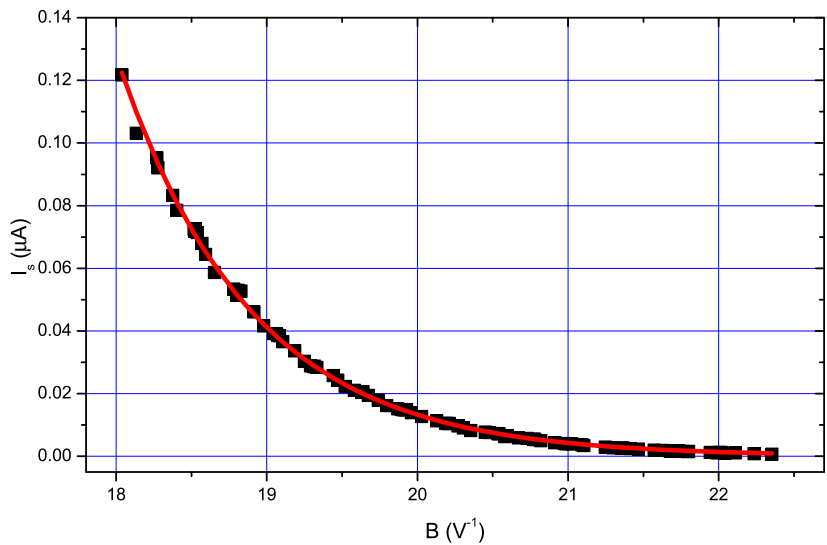}
\caption{$I_S$ as a function of $B$ for 100 \emph{I--V}  curves analysed with the iterative fitting procedure, described in the past subsection. The continuous curve represents the exponential \emph{fit} with the simplified dependence $I_S = I_A e^{-BE_G}$. The agreement is rather good and the best value of the parameters are $I_A = 68 \pm 2\, A$ and $E_G = 1.117 \pm 0.002\, eV$. The latter is in very close agreement with the tabulated value of the gap energy in Silicon at a temperature around 300 K.}
\label{fig7}
\end{figure}

The need to consider the factor $1/n$ in the exponential in eq. (\ref{Richa}) can be justified by our data, by tracing the graph of $I_S$ as a function of temperature, as determined by the exponential fit. The result is shown in Fig.~(\ref{fig7}) where a large set of curves \emph{I--V}, taken in a two hour period, are analysed. The weak power dependence on $T$ can be neglected with respect to the exponential and the fit of the data is performed with the simplified model:
$$ I_S = I_A\exp{\left(-E_G B\right)}$$
keeping the $B$ parameter defined above in the place of $e/nkT$ and expressing $E_G$ in $eV$. With this position the resulting parameters are $I_A = 68\pm 2\, A$ and $E_G = 1.117\pm 0.002\, eV$.

Other experiments using different data processing methods on Si \emph{p-n} junction assert a close agreement with the accepted \emph{band-gap} value of 1.12\, eV at 300 K. For example, \cite{Coll} report a value of 1.13 $\pm$ 0.02 eV, operating with the base-emitter junction of a Si bipolar transistor (2N3645 model), in the temperature interval $(-75, 25)\, ^\circ C$. It is not easy to find fabrication details of the 1N4148 diode and particularly information on the doping levels. Based exclusively on \cite{Orvis}, the 1N4148 is fabricated by growing a slightly doped ($<10^{16} \, cm^{-3}$)  $n-type$ Si micro-metric layer onto an heavily doped substrate $n^+$ at dopant concentration of $10^{19}\, cm^{-3}$. On the other side of $n$-layer an acceptor dopant is then diffused forming a $p$ layer with doping density of $10^{19}\, cm^{-3}$. With these levels of dopant a \emph{band-gap narrowing} of several tens of meV is predicted and observed \cite{Lanyon} and could conveniently be taken into account to adjust the expected value of $E_G$.

Our determination of $E_G$ is based on an approximate model but conceivably the $T^\delta$ factor in the $I_S(T)$ relation should be considered. It is easy to understand that as this factor increases with temperature, the simplified model underestimates diode current at higher temperatures and the characteristic constants of the exponential in function of $T$ lowers with respect to the value calculated without the exponential model:
$$I_S(T) = AT^\delta e^{-BE_G}$$
We take $\delta = 1.5$, which is the common choice for many different diodes, and the best fit gives the value of 1.040 $\pm$ 0.005 eV for $E_G$. The best fit $A$ value results $A = 2.5\pm 0.2\, mA/K^{1.5}$. The two parameters are strongly correlated so that an overestimation of the exponential characteristic constant involves an underestimation of the factor $A$ and the $E_G$ value is 80 meV lower than the one expected and with the disposable literature data \cite{Dhari}. This could  be explained in terms of a high doping level at $2$ to $3$ x $\, 10^{19}\, cm^{-3}$.

\subsection{Determination of the band gap from \emph{V--T} data}

For the evaluation of the energy band-gap, a different experimental procedure is more frequently followed (and also data analysis), consisting in the recording of $V$ at different temperatures at a fixed current $I$. It is possible to extract from our set of \emph{I--V}  curves the $V\, vs. \, T$ pairs at fixed diode current $\overline{I}$. The data processing consists in a linear interpolation of the value of $V$ between adjacent experimental pairs $(V_i,I_{i})$ and $(V_{i+1},I_{i+1})$ with $I_i<\overline{I}<I_{i+1}$.

\begin{figure}[!ht]
\centering
\includegraphics{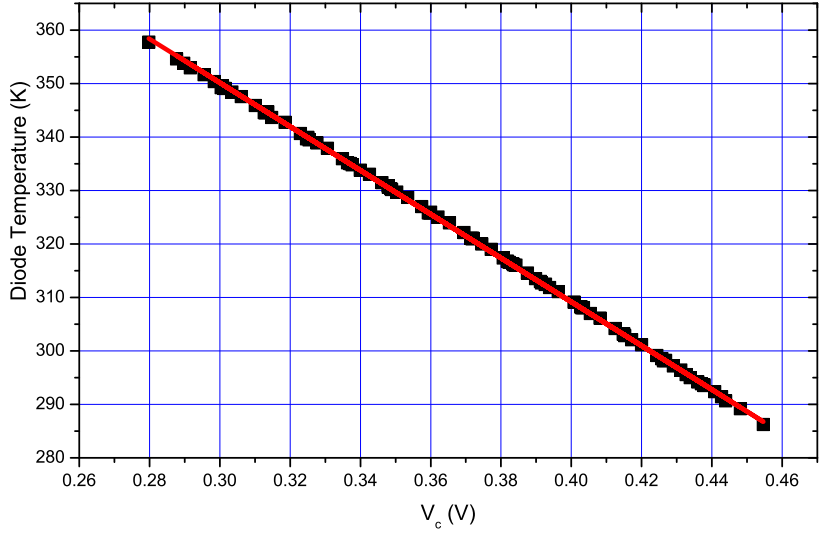}
\caption{$T$ as a function of $V$ as calculated interpolating (see text) the 100 \emph{I--V} curves analysed. The continuous curve represents the linear \emph{fit} $T = -aV + b$. $E_G =-b/a = 1.161\pm 0.002$ eV.}
\label{fig8}
\end{figure}

It is customary to choose $\overline{I}$ in a region where the Shockley characteristic curve can be  approximated by
\[
   \overline{I} = I_A\exp{\frac{V-eE_G}{nkT}}
\]
Taking the logarithm of both sides, a linear relation $T(V)$ is obtained:
\begin{equation}
\label{eq3}
T = -aV+b
\end{equation}
with
\[ a = \frac{1}{nk \log{\frac{I_A}{\overline{I}}}} \]
and
\[ b= \frac{eE_G}{nk\log{\frac{I_A}{\overline{I}}}} \]
so that with a simple linear fit the best values of $a$ and $b$ are calculated and:
\[ E_G (eV) = -\frac{b}{a} \]
where the possible dependencies of $n$ and $I_A$ cancel each other out. The final result of this analysis is shown in Fig.(\ref{fig8}) where the linearity of $T(V)$ is corroborated and the best fit values give $E_G=1.161\pm 0.002$ eV for $\overline{I} = 20 \mu A$. The values do not depend significantly on the choice of $\overline{I}$ in the region of moderate currents.

The overestimation of the expected value of $E_G$ is quite common in experiments driven at constant current and based on the analysis of the $T-V$ dependence. For example, \cite{Fischer} reports a value of 1.18 $\pm$ 0.02 eV at 300 K for the 1N4007 Si diode and \cite{Precker2} a value of 1.23 eV for the 1N4181. A good agreement with the accepted values of $E_G$ extrapolated at 0 K is reported by \cite{Canivez} which reports 1.19 $\pm$ 0.02 eV for the base emitter junction in a 2N2222 transistor and \cite{Kirkup} (1.165 $\pm$ 0.002 eV) in a 2N930 transistor. Finally, in \cite{Ocaya2} the authors analyze the $(T,V)$ data of a diode 1N4148, taken at different values of $I$ with a reciprocal procedure that extracts the temperature dependence of the saturation current. Assuming a pure exponential model for $I_S$ they find a value $E_G = 1.04\pm 0.02$ eV in very good agreement with the result found with our first method.

\section{Conclusion}

We have presented a detailed analysis of the direct current-voltage characteristic of a silicon diode 1N4148. A simple numerical method, based on the standard non linear fitting algorithms used by MatLab or Octave languages, is used with models of increasing complexity. The entire procedure of measurement and modelling can be easily automated integrating the data collection performed in a LabVIEW environment with suitable numerical packages.
The methodology used is relatively simple, and is suitable for an undergraduate science and engineering laboratory. The four parameters of the modified Shockley, represented in  eq.(\ref{3parhi}), are extracted and their values are found in quite good agreement with the literature.

The uncertainty analysis suggests that the accuracy of the approach presented here is comparable with other methods, and it is derived mainly from theoretical model than the statistical uncertainties, defect common to the majority of methods based on \emph{I--V} analysis and adopted in the current scientific and educational literature.

The advantages of the method presented here rely on the possibility to extract model parameters from data collected on a large span of $I$ values, and to easily compare simpler models, normally adopted in undergraduate laboratories.

Moreover, this analysis allows us to extract a value for $E_G$ within a few per cent in accordance with the accepted one, and the method is much more simple than the spectroscopic techniques which, furthermore, are also very difficult to be applied in a laboratory course.
Finally, we showed that different methods of analysis of the data bring results which do not coincide with the expected one and are sometimes inconsistent. From an educational point of view, this is valuable as it warns the student against the pitfalls hidden in the choice of the fit procedures even with the same model and the same data.

\end{document}